\documentclass[twocolumn,apl,amsmath,amssymb,showpacs]{revtex4-2}
\usepackage{epsf}      
\usepackage{graphicx}
\usepackage{color}
\usepackage{soul}
\usepackage{gensymb}
\usepackage{sidecap}
\usepackage{amsmath}
\usepackage{mathtools}
\begin{document}
\title{High temperature dielectric and impedance spectroscopy study of LaCo$_{0.7}$Nb$_{0.3}$O$_3$}
\author{Vikas N. Thakur}
\affiliation{Department of Physics, Indian Institute of Technology Delhi, Hauz Khas, New Delhi-110016, India}
\author{Ajay Kumar}
\affiliation{Department of Physics, Indian Institute of Technology Delhi, Hauz Khas, New Delhi-110016, India}
\author{Aakash Kaushik}
\affiliation{Department of Physics, Indian Institute of Technology Delhi, Hauz Khas, New Delhi-110016, India}
\author{Guru Dutt Gupt}
\affiliation{Department of Physics, Indian Institute of Technology Delhi, Hauz Khas, New Delhi-110016, India}
\author{Rajendra S. Dhaka}
\email{rsdhaka@physics.iitd.ac.in}
\affiliation{Department of Physics, Indian Institute of Technology Delhi, Hauz Khas, New Delhi-110016, India}

\date{\today}      

\begin{abstract}

We report the high temperature dielectric and {\it ac} impedance spectroscopy investigation of Nb substituted LaCo$_{0.7}$Nb$_{0.3}$O$_3$ polycrystalline sample. The maximum dielectric constant value was observed $\approx$1400 at around 400~K where the peak value shows a decreasing trend at higher temperatures and frequency. Similar variation was reflected in the dielectric loss (tan$\delta$) behavior with temperature, which shows the thermal activation of the charge carriers in the material. The analysis of high temperature impedance spectroscopy data shows the grain and grain boundary contributions by fitting the Nyquist plots to the equivalent circuit. From the analysis of the impedance and modulus spectra, it was possible to discern between the effects of overlapping grains, grain boundaries, and electrode interfaces. The relaxation time decreases with an increase in the temperature and the activation energy changes from 0.44~eV to 0.56~eV at around 400~K, which is due to involvement of thermal activation in the conduction of charge carriers. The conductivity is found to be increased with temperature for a given frequency, which again shows the semiconducting behavior. Whereas the conductivity increases with increase in frequency at lower temperatures. Also, the conductivity almost saturates with frequency at high temperatures.  
\end{abstract}

\maketitle
\section{\noindent ~Introduction}

The perovskite oxides, which have the general formula ABO$_3$, are therefore particularly desirable for technical devices such as solid oxide fuel cells (SOFCs), oxygen sensors, thermoelectric power generators, and other electrochemical devices \cite{42, 44, 45, 50, 51, 52}. The perovskite structured materials are most adaptable in the study of solid state chemistry by altering the chemical composition through aliovalent substitution and/or doping, where there are numerous opportunities to enhance and tune their physical properties, including conductivity (electronic and ionic), catalytic activity, thermal expansion control, thermoelectric, and magnetic behavior \cite{40, 41, 48}.  The cobaltites are a subclass of perovskite type oxides that have drawn sustained interest because of their superior electronic transport characteristics and mixed ionic-electronic conductivity, which make them a viable replacement for the current state-of-the-art SOFC cathode material. The rare-earth cobaltite LaCoO$_3$ shows exceptional catalytic activity and strong thermal stability \cite{SchmidtPRB09, Sudheendra}. Therefore, it has applications in a variety of disciplines, including energy production, catalysis, sensors and photocatalysis. By adjusting the temperature and pressure, as well as by making chemical replacements at the La/Co site, it shows peculiar magnetic and electronic phase transitions that can be adjusted \cite{AjayPRB20, AjayPRB2, AjayPRB22}. In this situation, the replacement of 3$d$ elements at the Co site in LaCoO$_3$ alters the spin state of neighbouring Co$^{3+}$ ions and exhibit a variety of anomalous behavior in the magnetic and transport characteristics \cite{17, 18, 19, 20, 21, 22, 23}. A metal-to-insulator transition was reported in LaCo$_{1-x}$Ni$_x$O$_3$ at a Ni substitution level of around 40\%, where the end members are non-magnetic insulator (LaCoO$_3$) and a paramagnetic metal (LaNiO$_3$) \cite{17,24}. The ferromagnetic/glassy nature and structural transition with Mn substitution have been observed in the magnetization and neutron diffraction investigation of LaCo$_{1-x}$Mn$_x$O$_3$, where the LaMnO$_3$ is an antiferromagnetic insulator and crystallizes in an orthorhombic structure \cite{23}. 

More interestingly, the high dielectric constant materials have gain renewed attention in the past several decades due to their extensive use in the various areas and devices like microwave filter, voltage control oscillator, dynamic access memory and telecommunication devices \cite{Vilar2005}. The earlier dielectric and ferroelectric material contain lead-based ceramic materials, which is very poisonous and harmful for the environment. So, the search of the suitable lead-free material with the desired dielectric properties is the major challenge of the present time \cite{SchmidtPRB09, Sudheendra}. In this context, with the substitution of Nb in LaCo$_{1-x}$Nb$_x$O$_3$, a structural change was reported from rhombohedral to orthorhombic/monoclinic and a decrease in electrical conductivity, which are addressed in terms of the spin state of the Co ion \cite{25, 39}. It was demonstrated that since the valence state of Co ion in LaCoO$_3$ is 3+, the charge balance with Nb substitution is achieved by reducing Co$^{3+}$ to Co$^{2+}$, as shown by the chemical formula LaCo$_{1-3x}^{3+}$Co$_{2x}^{2+}$Nb$_{x}^{5+}$O$_3^{2-}$; that is, Co is totally reduced to Co$^{2+}$ for $x=$  0.33 sample \cite{25, 39}. Additionally, the Nb$^{5+}$/Co$^{2+}$ ionic radius is larger than that of the Co$^{3+}$ ions, increasing the average radius at the Co location induces the lattice expansion \cite{25, 39}. In order to perform the impedance spectroscopy at high temperatures to understand charge conductivity in the material, this should either be semiconducting or insulating. This is because conversion of Co$^{3+}$ to Co$^{2+}$ due to Nb substitution at the Co-site decreases the conductivity of the material \cite{39, ShuklaJPCC19, ShuklaJPCC21}. Such a high impedance sample with the available dipoles for the polarization can be expected to be the suitable candidate for the interesting dielectric properties and change in conductivity \cite{NagaoPRB07}.

Therefore, in this article we investigate the dielectric properties and impedance spectroscopy of LaCo$_{0.7}$Nb$_{0.3}$O$_3$ at high temperatures as Nb doping at the Co-site is expected to increase the material's resistance and dielectric constant. The contributions of grain and grain boundary conductivity to overall conductivity can be distinguished using the well-established technique of impedance spectroscopy. The electric response is collected using the impedance method across a wide frequency range. The electric response may result from movements of the charge carriers in an electric field such as charge displacement, dipole reorientation, space charge generation, etc.

\section{\noindent ~Experimental}

The LaCo$_{0.7}$Nb$_{0.3}$O$_3$ was prepared by conventional ceramic method, as explained in earlier reported work \cite{39}. According to the composition of LaCo$_{0.7}$Nb$_{0.3}$O$_3$, the reagent-grade starting materials of lanthanum oxide (La$_2$O$_3$), cobalt oxide (Co$_3$O$_4$), and niobium pentoxide (Nb$_2$O$_5$) were mixed in a stoichiometric amount using mortar-pestle. The mixture was then calcined at 1100 $^o$C for 24 hrs. The ground powder was further pressed into a cylindrical shaped with the application of 80 MPa (diameter ~10 mm and thickness ~1 mm) followed by sintering at 1475$^o$C for 6 hrs and then slow cooling to room temperature at a rate of 5 $^o$C/min. After the sintering, the structural phase was analyzed by X-ray diffraction (XRD) (Cu K$_\alpha$ $\approx$ 1.5404 \AA) pattern in the Bragg's angle range from $20^o$ to $80^o$, Raman spectra from 300 to 900 cm$^{-1}$ wavenumbers, high resolution transmission electron microscopy (HR-TEM) and selected area (electron) diffraction (SAED) pattern. After confirmation of the phases and other structural parameters, the electrodes were deposited on ground disk surfaces using silver paste for the {\it ac} impedance spectroscopy measurements. The temperature ($T$) dependent impedance spectroscopy data were collected under vacuum using LCR meter (Manufacturer: HIOKI; model no. 3536) from 298~K to 588~K in the frequency range of 100~Hz--1~MHz.

\section{\noindent ~Results and discussion}

\begin{figure}
\includegraphics[width=3.5in,height=3.5in]{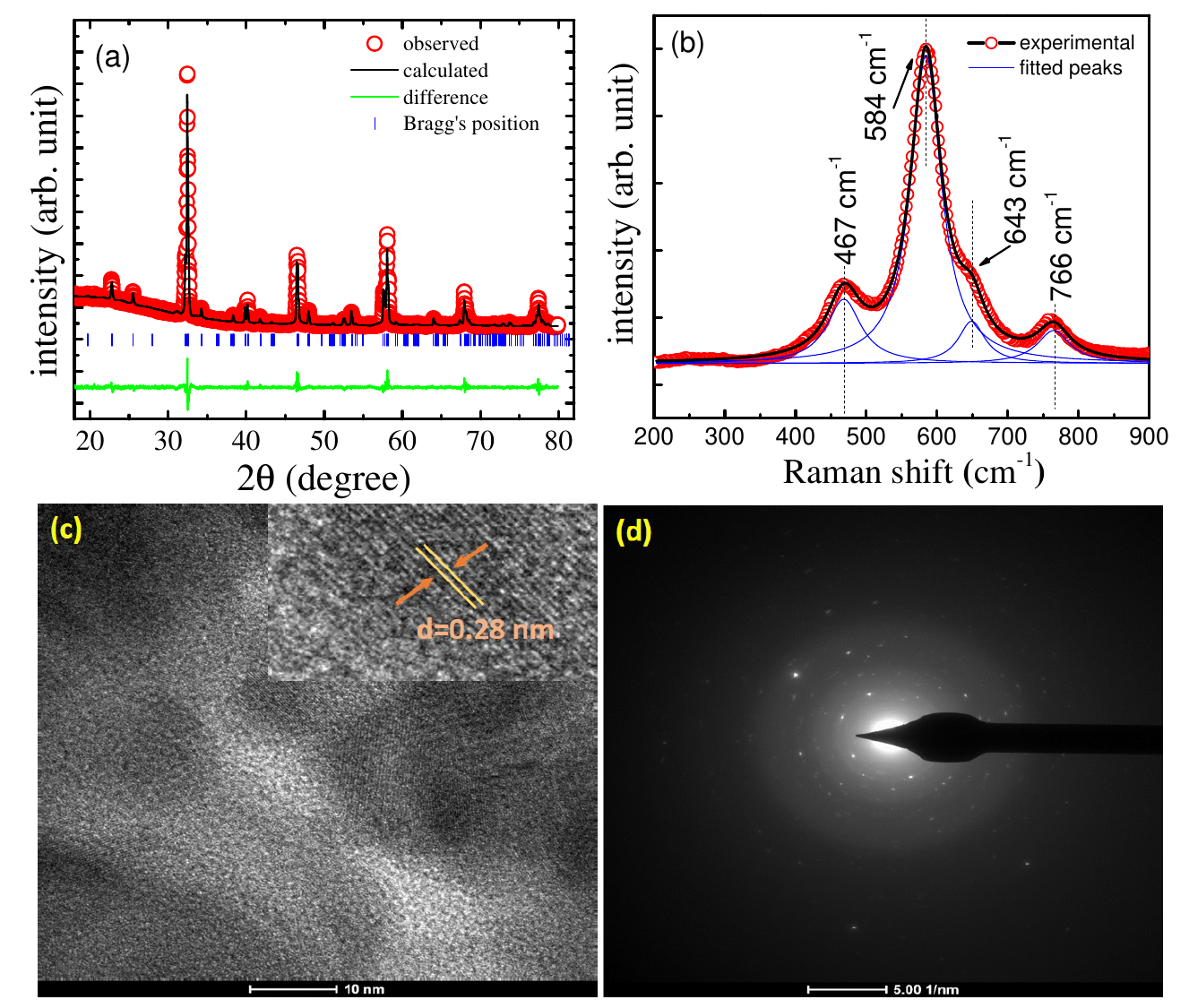}
\caption {(a) The XRD pattern in Bragg's angle range from $20^o$ to $80^o$, (b) the Raman spectra in the range from 300 to 900 cm$^{-1}$, (c) the HR-TEM micrograph with magnified image at the top right corner showing inter-planar spacing, and (d) the SAED pattern of the LaCo$_{0.7}$Nb$_{0.3}$O$_3$ sample.}
 \label{all_4}
\end{figure}

In Fig.~1(a), we show the Rietveld refined XRD pattern, which confirm the single phase polycrystalline nature of the sample and having the monoclinic crystal structure with the space group of $P2_1/n$. The lattice parameters obtained from the refinement are $a=$ 5.491~\AA, $b=$ 5.556 \AA, $c=$ 7.785 \AA, and $\beta=$ 89.99$^o$, which are found to be consistent with the one reported in literature \cite{ShuklaAIP}. The Raman spectrum in the range of 300--900 cm$^{-1}$, as shown in Fig.~1(b) with fitting with Lorentzian profile, shows four prominent active modes at around 467, 584, 643 and 766 cm$^{-1}$. Three peaks at 467, 584 and 766 cm$^{-1}$ are due to E$_g$ quadrupole, A$_{2g}$ breathing and E$_g$ phonon modes, respectively \cite{39}. The peak at 643 cm$^{-1}$ is due to increasing monoclinic phase in structure, which lead to lower its symmetry. Figures~1(c, d) show the TEM image and its high-resolution SAED pattern, respectively. The inset of TEM image shows the magnified image with calculated interplanar spacing of 0.28~nm and the SAED pattern confirms the polycrystalline nature of the material and its planes. The sample composition was confirmed by energy dispersive x-ray (not shown). 

The temperature dependent complex impedance ($Z$) and phase angle ($\theta$) are measured in the temperature range of 298--588 K using LCR meter under vacuum condition. The $Z$ and $\theta$ values are used to calculate complex dielectric constant ($\epsilon_r$), tangent loss (tan$\delta$), complex electric modulus ($M$), and {\it ac} conductivity using the following formulas given below:
\begin{equation}
Z = Z'+iZ''
\label{Debye}
\end{equation} 
\begin{equation}
M =i\omega C_0Z= M'+iM''
\end{equation} 
where, the $Z'$=$\lvert Z \rvert cos\theta$ and $Z''$=$\lvert Z \rvert sin\theta$ are real and imaginary part of complex impedance, and the $M'$ and $M''$ are real and imaginary part of complex electric modulus, respectively. The $\omega$ is angular frequency, and the capacitance in the vacuum is defined as $C_0$=$A$$\epsilon_0$/$d$, where $A$ is the area of electrode and $d$ is the separation between the electrodes. From the equations (1) and (2), we get
\begin{equation}
M'=\omega C_0Z'';M''=\omega C_0Z'
\end{equation} 
The dielectric constant can be given by inverse of electric modulus,
 \begin{equation}
\epsilon_r=1/M=\epsilon_r'+i\epsilon_r''
\end{equation}
where $\epsilon_r'$ and $\epsilon_r''$ are real and imaginary parts of $\epsilon_r$, respectivley.
Using the equations (2), (3) and (4), we get 
 \begin{equation}
\epsilon_r'=Z''/\omega C_0Z^2 ; \epsilon_r''=Z'/\omega C_0Z^2
\end{equation}
 \begin{equation}
tan\delta=\epsilon_r''/\epsilon_r'
\end{equation}

\begin{figure} 
\includegraphics[width=3.3in,height=4.6in]{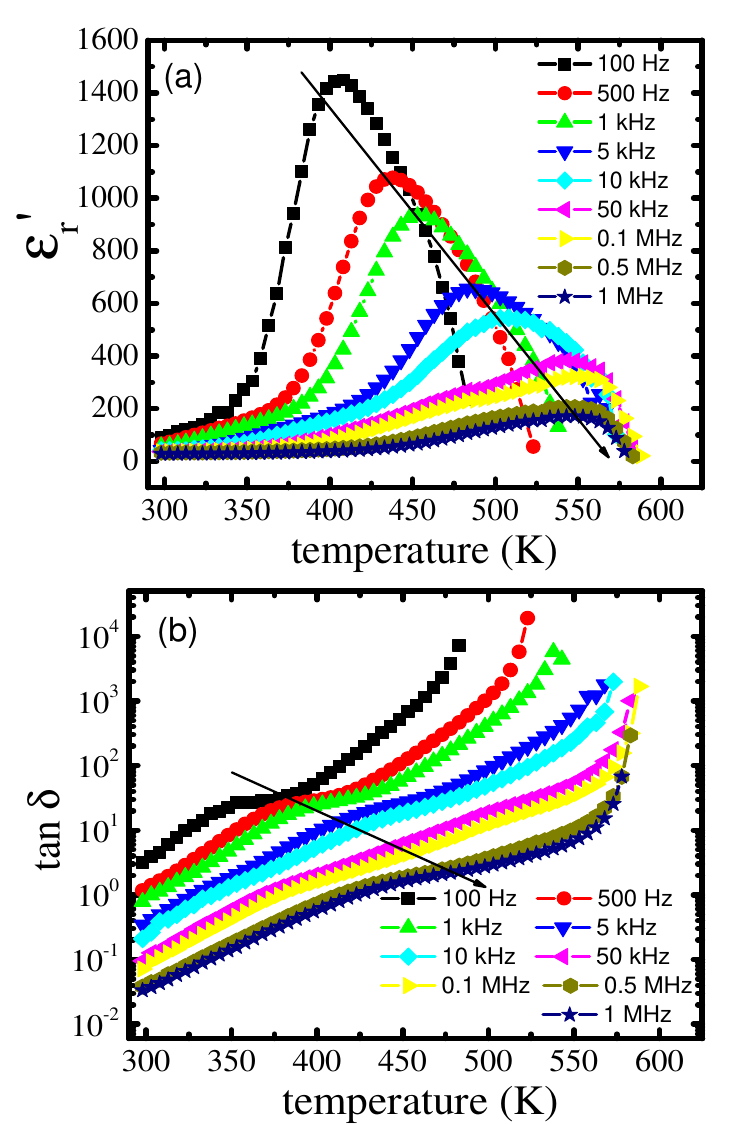}
\caption{(a) The real part of dielectric constant ($\epsilon_r'$) and (b) the dielectric loss (tan$\delta$) with varying temperature from 298--588~K in the frequency range of 100 Hz--1 MHz.} 
\label{fit_4}
\end{figure}

Now we present the calculated real part of dielectric constant and dielectric loss in Figs.~2(a, b), respectively with varying temperature from 298 K to 588 K. In Fig.~2(a), the peak shifts towards higher temperature and decreasing its amplitude with the frequency. It is customary to use thermal energy to illustrate how the dielectric constant changes with temperature \cite{26}. The mobility of charge carriers and, thus, their rate of hopping are improved by the additional thermal energy at higher temperatures, but at lower temperatures the thermal energy supplied is insufficient to increase the mobility of the charge carriers. As a result, higher temperature causes the dielectric polarization to grow, increasing the $\epsilon_r'$ with temperature up to a certain limit that is called the phase transition temperature where the polarization drops down to zero, therefore $\epsilon_r'$ starts decreasing with temperature. Similarly, the tan$\delta$ peaks, which are reflected from $\epsilon_r'$, are moving towards higher temperatures, as shown by the arrow in Fig.~2(b) at various frequencies. It should be noted that the tan$\delta$ curves have regular large peaks. Although, often this sort of behavior happens with a matching abrupt decrease in the $\epsilon_r'$, there is no comparable sharp drop in $\epsilon_r'$ curves, which calls attention to the underlying process \cite{27}. The tan$\delta$ peak maximum changes to a higher temperature with increasing frequency, which is a signature of the thermally triggered relaxation process. The dielectric relaxation losses frequently exhibit this characteristic \cite{28}. The {\it dc} conductivity input accounts for the majority of the quick increase in tan$\delta$ at higher tempaetures.

\begin{figure} 
\includegraphics[width=3.3in,height=4.6in]{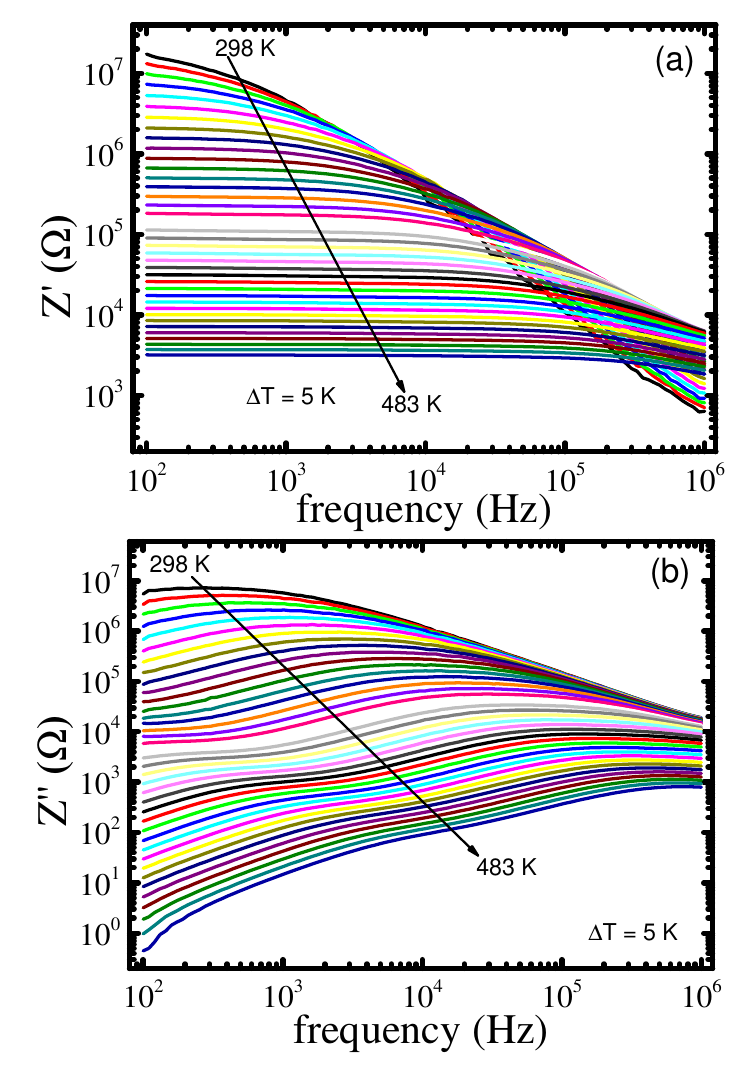}
\caption {(a) The real ($Z'$) and (b) imaginary ($Z''$) parts of impedance with varying frequency in the range of 100 Hz--1 MHz at various temperatures from 298 K to 483 K.} 
\label{fit_6}
\end{figure}

The sample's impedance and modulus spectra have been investigated in order to gain a better knowledge of the dielectric relaxation processes. Fig.~3(a) shows that the $Z'$ is decreasing with the temperature and frequency that explains the semiconducting behavior of the material. There are two regions in the impedance behavior: (1) frequency independent and (2) frequency dependent, which give rise to {\it dc} and {\it ac} conductivity of the material, respectively. The movement of charge carriers across extended distances, is possible by their successful hopping in which their neighboring charge carriers relax to their location, produces the {\it dc} conductivity. The limited or localized movement of charge carriers and their failed hopping in which they relax to their own places are indicators of dispersive {\it ac} conductivity. The relaxation maxima in the $Z''$ spectra shown in Fig.~3(b) offer details about the contribution of several parameters, including electrode interface effect and grain, grain boundary, etc. In the lower frequency band at room temperature, the  $Z''$ plot depicts a modest relaxation peak that can be attributed to the grain effect. The relaxation maxima migrate towards a higher frequency zone as the temperature rises, and their strength degrades. When the temperatures are elevated over 400~K, a weak peak in the low frequency range start to develop and progressively gets stronger. The large relaxation peaks in the higher frequency range followed by a very weak relaxation in the low-frequency region as a result of the electrode effect highlight the role of the grain and grain boundaries. Therefore, here we attempt to explain the peculiar reason behind the contribution from each grain, grain boundary, and electrode interface effect.

\begin{figure}[h] 
\includegraphics[width=3.8in]{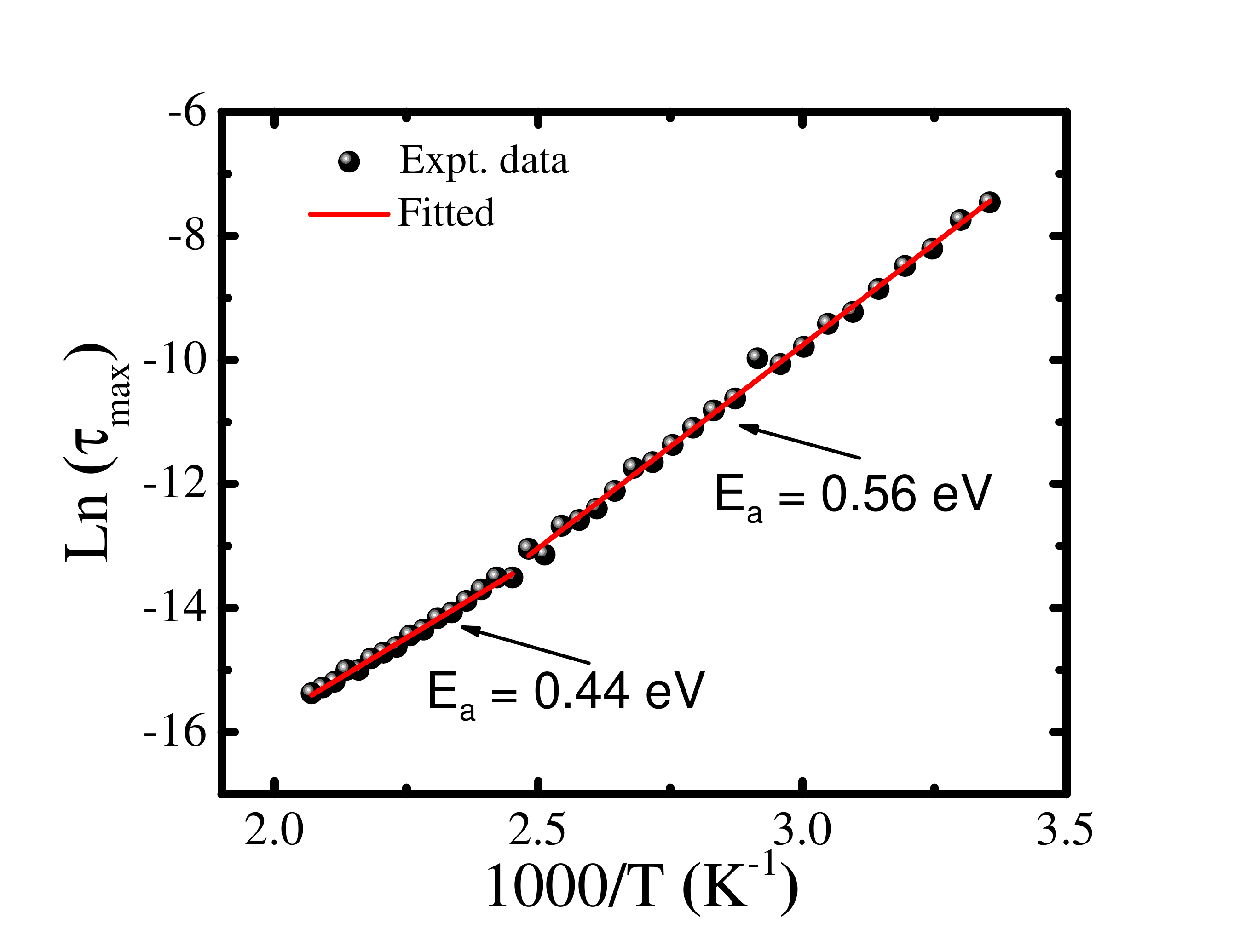}
\caption {The relaxation time with inverse of temperature fitted with Arrhenius model equation.} 
\label{fit_6}
\end{figure}

In order to understand the mechanism, we note that the peak position shifts to a higher frequency with increasing the temperature, which suggests that the relaxation rate has risen. The relationship 2$\pi \tau f_{max}$=1, where $f_{max}$ is the peak frequency of $Z''$ is used to compute the dielectric relaxation time ($\tau$). The temperature dependence of $\tau$ is presented in Fig.~4. We observe that the relaxation time reduces as temperature rises and we are able to fit its behavior with the Arrhenius relation: $\tau$=$\tau_0\exp^{-E_a/k_BT}$, where $E_a$ is the activation energy of the charge carriers, $k_B$ is the Boltzmann’s constant and $\tau_0$ is the pre-exponential factor of the relaxation time. Using the plot given in Fig.~4, the calculated $E_a$ values are found to be 0.56~eV and 0.44~eV in lower and higher temperatures, respectively. It can be seen that the activation energy decreases by 0.12 eV above 400~K and this reduction is because above this temperature the charge carrier is already thermally activated, that is why the carriers need less activation energy than below 400~K.

Figures~5(a--d) display the Nyquist plots ($Z''$ vs. $Z'$) in different temperature ranges in 5~K step, which are fitted with the analogous circuit shown in the inset of Fig.~5(a). The data are often described by an ideal equivalent electrical circuit consisting of resistance ($R$) and capacitance ($C$) in order to examine the impedance behavior and that creates a link between microstructure and electrical attributes. The resistor in the circuit represents the conductive path and usually considers the bulk conductivity of the synthesized material. Similarly, the capacitor in the circuit associates with the space charge polarization regions at the electrode interface. The majority of polycrystalline materials exhibit both grain and grain-boundary impedances. We demonstrate that the comparable circuit built using the brick-layer model, as shown in the inset of Fig.~5(a), better fits the impedance data. This circuit consists of a series array of two sub-circuits, one of which represents the grain effects and the other grain boundaries.

\begin{figure}[h] 
\includegraphics[width=3.4in,height=3.2in]{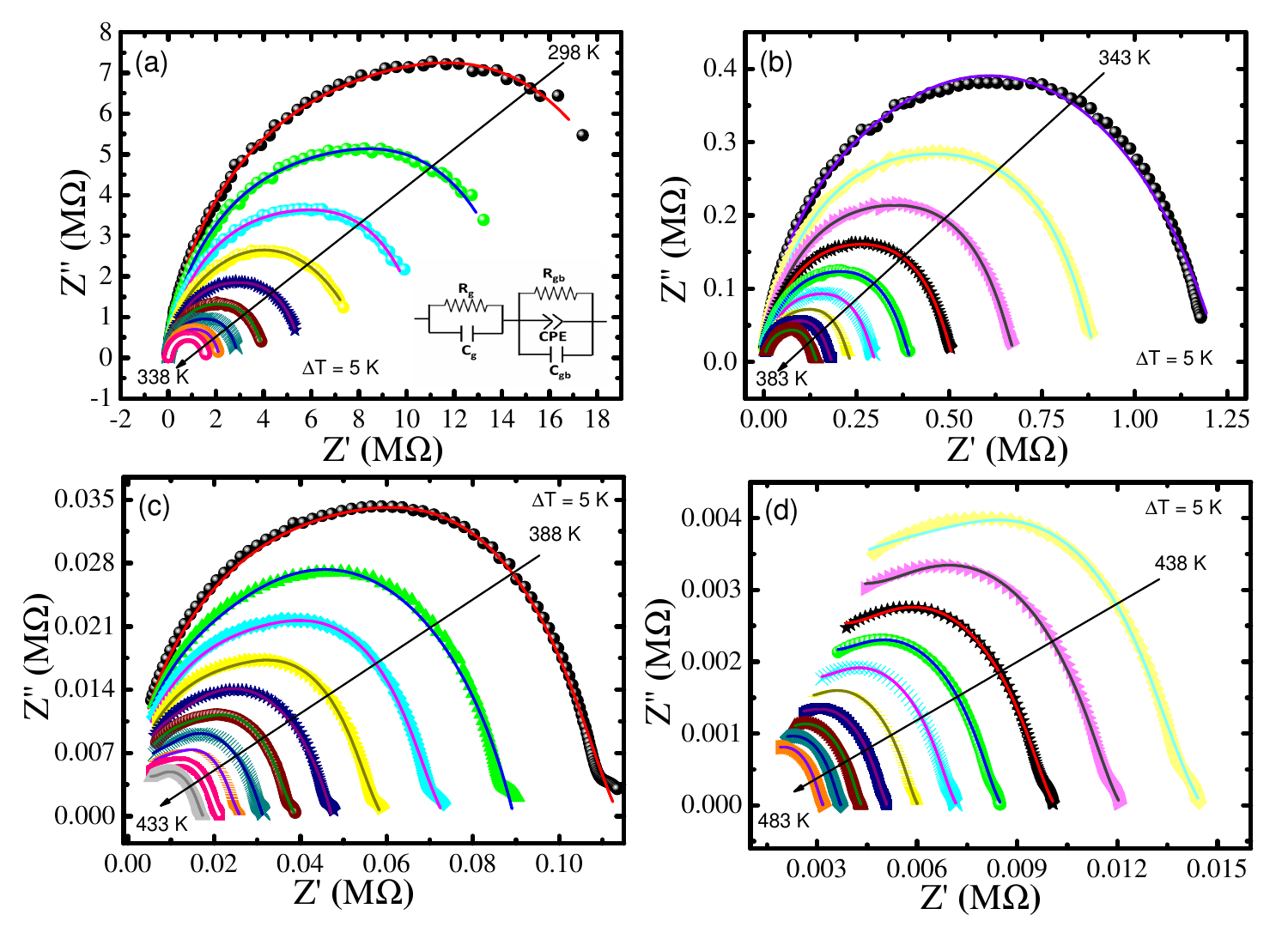}
\caption {The Nyquist plots in the range of (a) 303 – 338 K, (b) 343 – 383 K, (c) 388 K – 433 K and (d) 438 K – 483 K fitted with electric circuit shown in the inset of (a).} 
\label{short_4}
\end{figure}

Therefore, we denote the ($R_g$, $R_{gb}$) and ($C_g$, $C_{gb}$) as the resistances and capacitances for the individual grains and their respective grain boundaries, respectively. 
\begin{figure}[h] 
\includegraphics[width=3.3in]{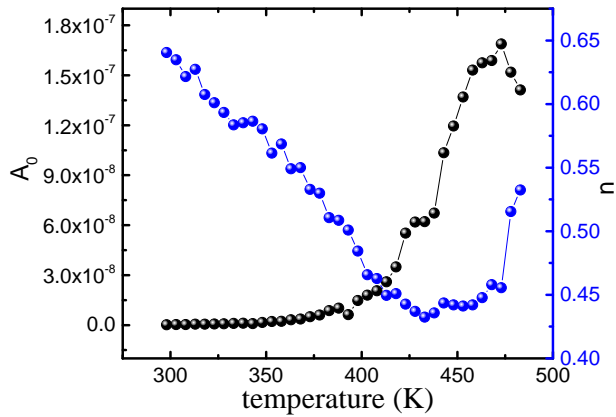}
\caption {The plot of $A_0$ and $n$ with varying temperature in the range of 298 – 483 K.} 
\end{figure}
The $CPE$ stands for constant phase element, which denotes non-Debye-type relaxation behavior \cite{29} as all the dipoles are not relaxed with the same relaxation durations according to the dielectric behavior. The plots shown in Figs.~5(a-d) also demonstrate that the impedance decreases with temperature more quickly (from $M\Omega$ to $k\Omega$). The $CPE$ impedance is represented by the expression 1/$A_0 (i\omega)^n$, where $A_0$ and $n$ are temperature-dependent factors only, $A_0$ limits the dispersion's magnitude, and $0<n<1$. For an ideal resistor and capacitor, the parameter of $n$ is equal to 0 and 1, respectively \cite{30}. The Nyquist plots were fitted using ZSIMPWIN 3.2.1 software under the assumption of the aforementioned equivalent circuit. The semicircular arc's radius, which determines the material's resistance, providing the insights into numerous conduction mechanisms. A semicircular arc forms at ambient temperature and lasts up to 338 K, spanning both the high- and low-frequency regimes. According to our assumptions, the grain contribution is significantly reduced by the electrode contact and grain boundary effects from ambient temperature to 338~K \cite{ThongbaiJPCM08}.
  
As we can see that the grain boundary contribution increases gradually with increase in the temperature, and it emerges at 343~K to 383~K in exactly the same amounts for both the grain and the grain boundary. Above 383~K, the lower frequency zone predicts an increase in the conduction process at the electrode interface, 
 \begin{figure}[h] 
\includegraphics[width=3.4in,height=3.25in]{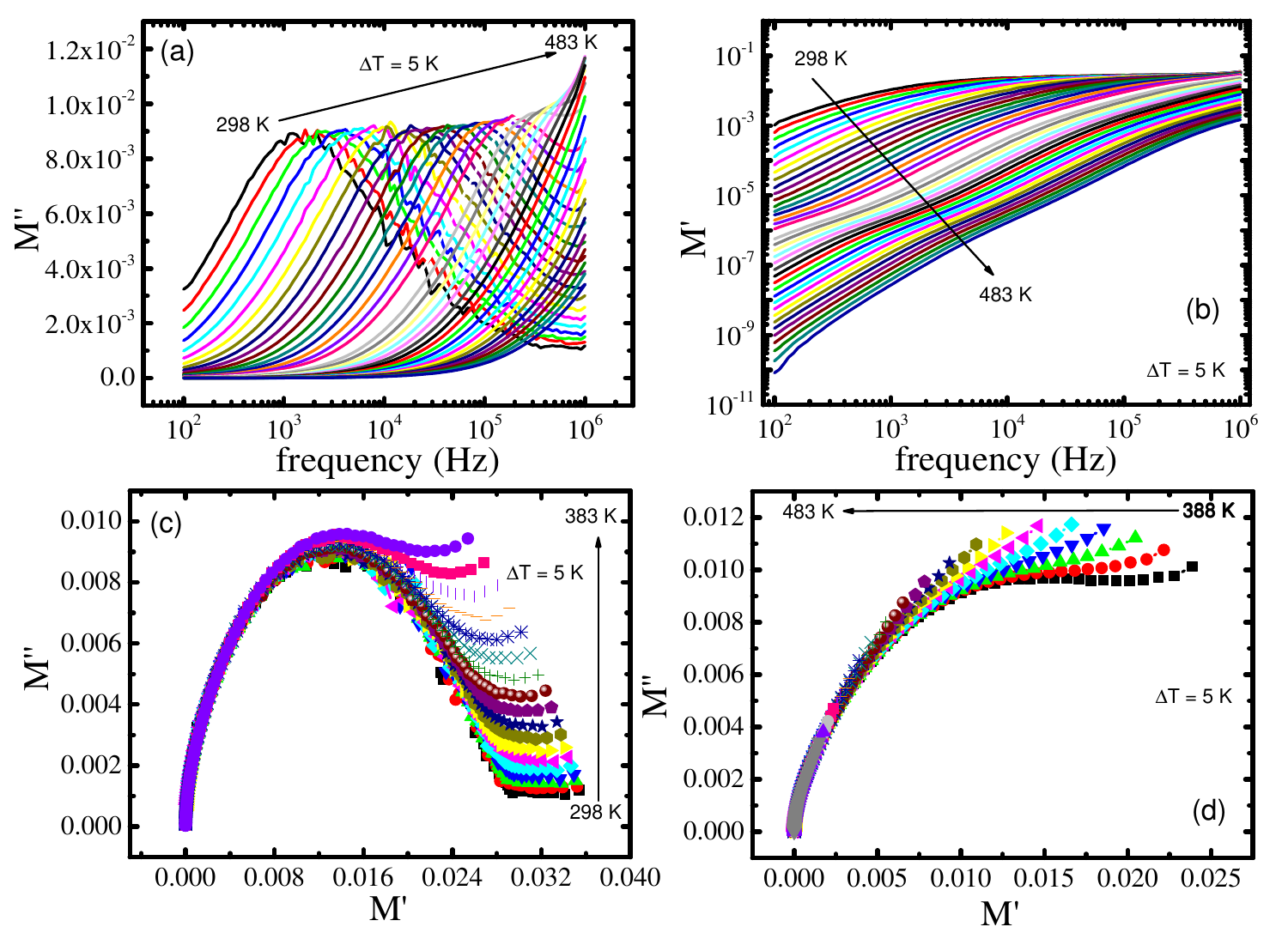}
\caption {(a) The imaginary ($M''$) and (b) real ($M'$) parts of the electric modulus with varying frequency in the range of 100 Hz--1 MHz at various temperatures from 298 to 483 K. The Cole-Cole plot of modulus in the temperature range of (c) 298 K – 383 K and (d) 388 K – 483 K.} 
\end{figure}
whereas the high-frequency region suggests that the behavior of grains and grain boundaries should become more uniform. The fitted values of $n$ of $CPE$ are in the range of 0.44--0.65 and initially declines, then increases at higher temperatures, as plotted in the Fig.~6 on right scale. The fitted value of $A_0$ are plotted on the left scale of Fig.~6, which are found to be almost constant up to 350 K, then increases steadily up to 403~K, and after that it climbs quickly when the temperature increase up to around 470~K. This is possibly due to the interference between grain boundary effect and electrode contact conduction in the material \cite{RaniCI17, NobreJAP03}. 

It is important to note here that in contrast to the grain boundaries and the electrode effect, the fictitious impedance loss spectra show a weak grain impact that can be challenging to analyze. To magnify and manifest the grain impacts, the electric modulus spectra emphasize and accentuate lower capacitance values. Therefore, in Figs.~7(a, b) we show the frequency dependent real and imaginary parts of modulus spectra over a broad frequency range and between 298~K and 483~K temperatures. 
\begin{figure}[h] 
\includegraphics[width=3.3in]{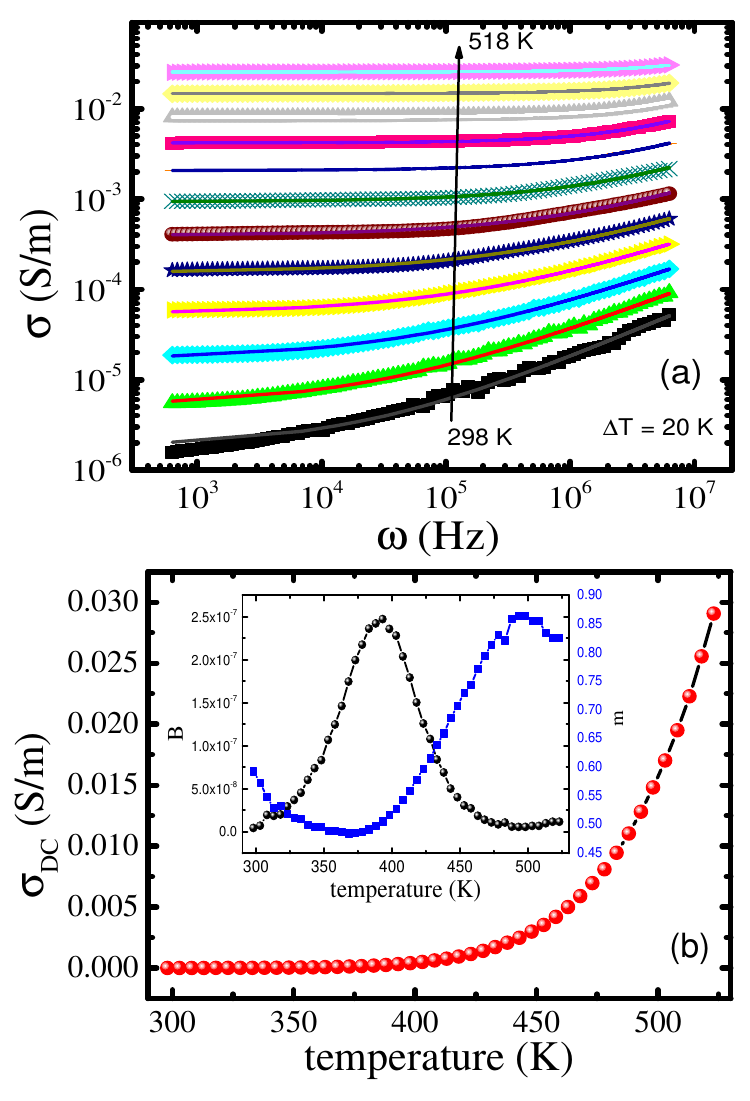}
\caption {(a) The {\it ac} conductivity variation with frequency measured at various temperatures in the range of 298--518~K. (b) The Jonscher power law extracted {\it dc} conductivity in the temperature range of 298--518~K. The variation in parameters $B$ and $m$ with temperature is shown in the inset of (b).} 
\end{figure}
At around room temperature, the real modulus continually disperses with frequency, and that as the temperature rises, the dispersion progressively reaches to the saturation. The imaginary modulus spectra show two relaxation peaks that start to develop at ambient temperature and last until 383~K. One of the relaxation peaks disappears above 383~K. We have drawn a Cole-Cole modulus plots in Figs.~7(c, d) to comprehend the process underlying this relaxation \cite{TriyonoRSC20}. The Cole-Cole plots of impedance typically combine the relaxation of the electrode interface, grain, and grain boundary into a single semicircle. The modulus spectra that can resolve into a different semicircular arc are illustrated to help distinguish each relaxing step. One semicircular arc can be seen in the $M'$ vs. $M''$ spectra [ see Figs.~7(c, d)], which indicates that one of the relaxation processes has been severely inhibited by the surface polarization effect. A minor grain contribution is visible when a little arc forms at the end of the $M'$ axis; this arc eventually vanishes at higher temperatures. On the other hand, the bigger semicircular arc that results from grain boundary relaxation and the shortened form of this arc with rising the temperature are examples of how the grain boundary contribution eventually diminishes. Above 383~K, the contribution of the electrode ceramic contact becomes more significant, and the Cole-Cole plot of impedance data provides good support to this approach. The real and imaginary modulus spectra show that the high temperature electrode polarization revealed by impedance spectroscopy is severely inhibited in this situation \cite{31,32}.

Finally, in Fig.~8(a) we the fitted total conductivity ($\sigma$) data with varying the angular frequency ($\omega$) in the temperature range of 298--518~K using Jonscher Power Law equation: $\sigma=\sigma_{dc}+B\omega^m$, where $\sigma_{dc}$ is the {\it dc} conductivity, $B\omega^m$ is the {\it ac} conductivity ($\sigma_{ac}$) where $B$ and $m$ are polarization strength and degrees of interaction of charge carriers, respectively and the temperature dependent parameters. We observe that the $\sigma$ values increases with increase in temperature and gradual increase in conductivity is also clearly visible with frequency, which shows their strong frequency dependence. The conductivity increases towards higher frequency at lower temperatures but moving towards saturation at higher temperatures, which presents the threshold conductance of the material. Figure~8(b) shows the $\sigma_{dc}$ plot with varying the temperature where the variations in $B$ and $m$ with temperature are shown in the inset. We found the $\sigma_{dc}$ constant initially but above around 400~K, it starts gradually increasing which shows the semiconducting nature of the material. The $m$ values decrease at lower temperatures then starts increasing near 400~K, which shows the overlapping large polaron tunneling (OLPT) conduction mechanism \cite{33, 34}. The $m$ value rises linearly at higher temperatures, which explains the non-overlapping small polaron hopping (NSPT) behaviour \cite{35}. The $m$ value reaches around 0.85 and saturates around 500~K temperature, which can be understood by quantum mechanical tunneling (QMT) \cite{36}. However, above 500~K, the $m$ starts decreasing slightly and then saturates approximately, which shows the correlated barrier hopping (CBH) \cite{37, 38, TaherAPA15}. Similarly, the $B$ values first increase with temperature and peaks near 400~K, that means the polarization strength is maximum around this temperature, then it starts decreasing exponentially, which is the reflection of the transition, and found to be consistent with the dielectric behavior as shown in Fig.~2(a). \\

\section{\noindent ~Conclusions}

The polycrystalline LaCo$_{0.7}$Nb$_{0.3}$O$_3$ sample was prepared using solid state reaction method and its characterization was performed using XRD, Raman spectroscopy, and HR-TEM. The peak value of the dielectric constant decreases with a monotonic shift towards higher temperature from 298~K to 583~K. Similarly, the dielectric loss increases at higher temperatures rapidly because its dielectric constant is very low and hence becomes more conducting. This is consistent as the conductivity of the material increases with temperature. The impedance spectroscopy study suggests a decrease in magnitude of impedance with temperature, which represents the semiconducting nature of the material. The peaks shift towards higher frequency in the imaginary part and decreases with temperature, which explain the relaxation behavior of charge carriers. The same behavior is also observed in relaxation time variation with temperature. The {\it ac} conductivity increases with frequency at lower temperatures but, it saturates at higher temperatures, which again shows the threshold of the conductance in the material. Similarly, the {\it dc} conductivity increases with temperature, which represents the semiconducting nature of the sample.

%\section{\noindent Author contributions:}

%All the authors have contributed to this work.

%\section{\noindent ~Declaration of competing interest}
%The authors declare that they have no known competing financial interests or personal relationships that could have appeared to influence the work reported in this paper. 

%\section{\noindent~DATA AVAILABILITY}

%The data that support the findings of this study are available from the corresponding author upon reasonable request.
	
\section{\noindent ~Acknowledgments}

V.N.T. and Ajay Kumar thank DST (project: DST/TMD/MECSP/2K17/07) and UGC, respectively, and Aakash Kaushik and G.D.G. thank MHRD for the fellowship. Authors also thank the IIT Delhi for providing characterization facilities: XRD and Raman at department of physics and HR-TEM at CRF. RSD acknowledges the financial support from SERB--DST through the core research grant (file no: CRG/2020/003436).


\begin{thebibliography}{99}

\bibitem{42} J. Chakrabartty, C. Harnagea, M. Celikin, F. Rosei, and R. Nechache, Improved photovoltaic performance from inorganic perovskite oxide thin films with mixed crystal phases, Nat. Photonics. {\bf 12} (2018) 271.

\bibitem{44} S. Tao and J. T. S. Irvine, A redox-stable efficient anode for solid-oxide fuel cells, Nat. Mater. {\bf 2} (2003) 320.

\bibitem{45} J. Sunarsoa, S. S. Hashima, N. Zhub, W. Zhou, Perovskite oxides applications in high temperature oxygen separation, solid oxide fuel cell and membrane reactor: a review, Prog. Energy Combust. Sci. {\bf 61} (2017) 57.

\bibitem{50} W. Yu, F. Li, L. Yu, M. R. Niazi, Y. Zou, D. Corzo, A. Basu, C. Ma, S. Dey, M. L. Tietze, U. Buttner, X. Wang, Z. Wang, M. N. Hedhili, C. Guo, T. Wu, and A. Amassian, Single crystal hybrid perovskite field-effect transistors, Nat. Commun. {\bf 9} (2018) 5354.

\bibitem{51} B. Jeong, L. Veith, T. J. A. M. Smolders, M. J. Wolf, and K. Asadi, Room‐temperature halide perovskite field‐effect transistors by ion transport mitigation, Adv. Mater. {\bf 33} (2021) 2100486.

\bibitem{52} C. C. Homes, T. Vogt, S. M. Shapiro, S. Wakimoto, and A. P. Ramirez, Optical response of high-dielectric-constant perovskite-related oxide, Science {\bf 293} (2001) 673.

\bibitem{40} A. Chainani, M. Mathew, and D. D. Sharma, Electron-spectroscopy study of the semiconductor-metal transition in La$_{1-x}$Sr$_x$CoO$_3$,
Phys. Rev. B {\bf 46} (1992) 9976.

\bibitem{41} Q. Sun, J. Wang, W. -J. Yin, and Y. Yan, Bandgap engineering of stable lead-free oxide double perovskites for photovoltaics, Adv. Mater {\bf 30} (2018) 1705901.

\bibitem{48} C. Gauvin-Ndiaye, A. -M. S. Tremblay, and R. Nourafkan, Electronic and magnetic properties of the double perovskites La$_2$MnRuO$_6$ and LaAMnFeO$_6$ (A= Ba, Sr, Ca) and their potential for magnetic refrigeration, Phys. Rev. B {\bf 99} (2019) 125110.


\bibitem{SchmidtPRB09} R. Schmidt, J. Wu, C. Leighton, and I. Terry, Dielectric response to the low-temperature magnetic defect structure and spin state transition
in polycrystalline LaCoO$_3$, Phys. Rev. B {\bf 79} (2009) 125105.

\bibitem{Sudheendra} L. Sudheendra, M. Seikh, A. R. Raju, C. Narayana, and C. N. R. Rao, Dielectric properties of rare earth cobaltates, LnCoO$_3$ (Ln=La, Pr, Nd), across the spin-state transition, Ferroelectrics {\bf 306} (2010) 227.

\bibitem{AjayPRB20} A. Kumar, and R. S. Dhaka, Unraveling the magnetic interactions and spin state in insulating Sr$_{2-x}$La$_x$CoNbO$_6$, Phys. Rev. B {\bf 101} (2020) 094434.

\bibitem{AjayPRB2} A. Kumar, B. Schwarz, H. Ehrenberg, and R. S. Dhaka, Evidence of discrete energy states and cluster-glass behavior of Sr$_{2-x}$La$_x$CoNbO$_6$, Phys. Rev. B {\bf 102}  (2020) 184414.

\bibitem{AjayPRB22} A. Kumar, R. Shukla, R. Kumar, R. J. Choudhary, S. N. Jha, and R. S. Dhaka, Electronic and local structure investigation of Sr$_{2-x}$La$_x$CoNbO$_6$ using near-edge and extended x-ray absorption fine structures, Phys. Rev. B {\bf 105} (2022) 245155.

\bibitem{17}D. Hammer, J. Wu, C. Leighton, Metal-insulator transition, giant negative magnetoresistance, and ferromagnetism in ${\mathrm{LaCo}}_{1\ensuremath{-}y}{\mathrm{Ni}}_{y}{\mathrm{O}}_{3}$, Phys. Rev. B {\bf 69} (2004) 134407. 

\bibitem{18}K. Tomiyasu, Y. Kubota, S. Shimomura, M. Onodera, S. -I. Koyama, T. Nojima, S. Ishihara, H. Nakao, Y. Murakami, Spin-state responses to light impurity substitution in low-spin perovskite LaCoO$_{3}$, Phys. Rev. B {\bf 87} (2013) 224409. 

\bibitem{19}M. Viswanathan, P. S. A. Kumar, Observation of reentrant spin glass behavior in ${\text{LaCo}}_{0.5}{\text{Ni}}_{0.5}{\text{O}}_{3}$, Phys. Rev. B {\bf 80} (2009) 012410. 
 
\bibitem{20}V. Kumar, R. Kumar, D. K. Shukla, S. K. Arora, I. V. Shvets, K. Singh, R. Kumar, Spin states and glassy magnetism in $\mathrm{LaCo}_{1-x} \mathrm{Ni}_{x} \mathrm{O}_{3}(0 \leq x \leq 0.5)$, Materials Chemistry and Physics {\bf 147} (2014) 617. 

\bibitem{21}V. Kumar, R. Kumar, D. K. Shukla, S. Gautam, K. Hwa Chae, R. Kumar, Electronic structure and electrical transport properties of $\mathrm{LaCo}_{1-x} \mathrm{Ni}_{x} \mathrm{O}_{3}(0 \leq x \leq 0.5)$, J. Appl. Phys. {\bf 114} (2013) 073704. 

\bibitem{22}N. E. Rajeevan, V. Kumar, R. Kumar, R. Kumar, and S. D. Kaushik, Neutron diffraction studies of magnetic ordering in Ni-doped LaCoO$_3$, Journal of Magnetism and Magnetic Materials {\bf 393} (2015) 394. 

\bibitem{23}C. L. Bull, H. Y. Playford, K. S. Knight, G. B. G. Stenning, and M. G. Tucker, Magnetic and structural phase diagram of the solid solution ${\text{LaCo}}_{x}{\text{Mn}}_{1\ensuremath{-}x}{\text{O}}_{3}$, Phys. Rev. B {\bf 94} (2016) 014102. 

\bibitem{24}K. P. Rajeev, and A. K. Raychaudhuri, Quantum corrections to the conductivity in a perovskite oxide: a low-temperature study of ${\mathrm{LaNi}}_{1\mathrm{\ensuremath{-}}\mathit{x}}$${\mathrm{Co}}_{\mathit{x}}$${\mathrm{O}}_{3}$ (0\ensuremath{\le}x\ensuremath{\le}0.75), Phys. Rev. B {\bf 46} (1992) 1309. 

\bibitem{Vilar2005} S. Yanez-Vilar, A. Castro-Couceiro, B. Rivas-Murias, A. Fondado, J. Mira, J. Rivas, and M. A. Senarıs-Rodrıguez, Study of the dielectric properties of the perovskite LaMn$_{0.5}$Co$_{0.5}$O$_{3-\delta}$, Z. Anorg. Allg. Chem. {\bf 631} (2005) 2265 .

\bibitem{25}V. Oygarden, H.L. Lein, T. Grande, Structure, thermal expansion and electrical conductivity of Nb-substituted LaCoO$_3$, Journal of Solid State Chemistry {\bf 192} (2012) 246. 

\bibitem{39}R. Shukla, R. S. Dhaka, Anomalous magnetic and spin glass behavior in Nb substituted LaCo$_{1-x}$Nb$_x$O$_3$, Phys. Rev. B {\bf 97} (2018) 024430. 

\bibitem{ShuklaJPCC19} R. Shukla, A. Jain, M. Miryala, M. Murakami, K. Ueno, S. M. Yusuf, and R. S. Dhaka, Spin dynamics and unconventional magnetism in insulating La$_{(1-2x)}$Sr$_{2x}$Co$_{(1-x)}$Nb$_x$O$_3$, Journal of Physical Chemistry C {\bf 123} (2019) 22457.

\bibitem{ShuklaJPCC21} R. Shukla, A. Kumar, R. Kumar, S. N. Jha, and R. S. Dhaka, X-ray absorption spectroscopy study of La$_{1-y}$Sr$_y$Co$_{1-x}$Nb$_x$O$_3$, Journal of Physical Chemistry C {\bf 125} (2021) 10130.

\bibitem{NagaoPRB07} Y. Nagao and I. Terasaki, Dielectric constant and ac conductivity of the layered cobalt oxide Bi$_2$Sr$_2$CoO$_{6+}$: a possible metal-dielectric composite made by self-organization of Co$^{2+}$ and Co$^{3+}$ ions, Phys. Rev. B {\bf 76} (2007) 144203.

\bibitem{ShuklaAIP}R. Shukla, and R. S. Dhaka, Physical properties of Sr and Nb substituted LaCoO$_3$, AIP Conf. Proc. {\bf 2265} (2020) 030586.

\bibitem{26}Y. Song, Q. Sun, Y. Lu, X. Liu, F. Wang, Low-temperature sintering and enhanced thermoelectric properties of LaCoO$_3$ ceramics with B$_2$O$_3$–CuO addition, Journal of Alloys and Compounds {\bf 536} (2012) 150. 

\bibitem{27}D. C. Sinclair, T. B. Adams, F. D. Morrison, and A. R. West, CaCu$_3$Ti$_4$O$_12$: one-step internal barrier layer capacitor, Appl. Phys. Lett {\bf 80}  (2002) 2153. 

\bibitem{28}I. S. Zheludev, Physics of crystalline dielectrics, electrical properties, Plenum Press, New York, London (1971).

\bibitem{29}R. M. Hill, and L. A. Dissado, Debye and non-Debye relaxation, J. Phys. C: Solid State Phys {\bf 18} (1985) 3829 . 

\bibitem{30}A. ur Rahman, M. A. Rafiq, K. Maaz, S. Karim, S. Oh Cho, and M. M. Hasan, Temperature induced delocalization of charge carriers and metallic phase in Co$_{0.6}$Sn$_{0.4}$Fe$_2$O$_4$ nanoparticles, J. Appl. Phys. {\bf 112} (2012) 063718. 

\bibitem{ThongbaiJPCM08} P. Thongbai, S. Tangwancharoen, T. Yamwong and S. Maensiri, Dielectric relaxation and dielectric response mechanism in (Li, Ti)-doped NiO ceramics, J. Phys. Condens. Matter {\bf 20} (2008) 395227.

\bibitem{RaniCI17} S. Rani, N. Ahlawat, R. Punia, K. M. Sangwan and S. A. F. Rani, Dielectric relaxation and conduction mechanism of complex perovskite Ca$_{0.90}$Sr$_{0.10}$Cu$_3$Ti$_{3.95}$Zn$_{0.05}$O$_{12}$ ceramic, Ceram. Int. {\bf 44} (2017) 5996.

\bibitem{NobreJAP03} M. A. L. Nobre and S. Lanfredi, Ferroelectric state analysis in grain boundary of Na$_{0.85}$Li$_{0.15}$NbO$_3$ ceramic, J. Appl. Phys. {\bf 93} (2003) 5557.

\bibitem{TriyonoRSC20} D. Triyono, S.N. Fitria and U. Hanifah, Dielectric analysis and electrical conduction mechanism of La$_{1-x}$Bi$_x$FeO$_3$ ceramics, RSC Adv. {\bf 10}  (2020) 18323.

\bibitem{31}N. Sivakumar, A. Narayanasamy, C. N. Chinnasamy, and B. Jeyadevan, Influence of thermal annealing on the dielectric properties and electrical relaxation behaviour in nanostructured CoFe$_2$O$_4$ ferrite, J. Phys.: Condens. Matter {\bf 19} (2007) 386201. 

\bibitem{32}A. Srivastava, A. Garg, and F. D. Morrison, Impedance spectroscopy studies on polycrystalline BiFeO$_3$ thin films on Pt/Si substrates, J. Appl. Phys. {\bf 105}  (2009) 054103. 

\bibitem{33}A. Kahouli, A. Sylvestre, F. Jomni, B. Yangui, and J. Legrand, Experimental and theoretical study of ac electrical conduction mechanisms of semicrystalline parylene C thin films, J. Phys. Chem. A {\bf 116} (2012) 1051. 

\bibitem{34}A. R. Long, Frequency-dependent loss in amorphous semiconductors, Advances in Physics {\bf 31} (1982) 553. 

\bibitem{35}A. Ghosh, Transport properties of vanadium germanate glassy semiconductors, Phys. Rev. B {\bf 42} (1990) 5665. 

\bibitem{36}A. Ghosh, Frequency-dependent conductivity in bismuth-vanadate glassy semiconductors, Phys. Rev. B {\bf 41} (1990) 1479. 

\bibitem{37}Y. B. Taher, A. Oueslati, K. Khirouni, and M. Gargouri, Impedance spectroscopy and conduction mechanism of LiAlP$_2$O$_7$ material, Materials Research Bulletin {\bf 78} (2016) 148. 

\bibitem{38}S. Mollah, K. K. Som, K. Bose, and B. K. Chaudhuri, AC conductivity in $\mathrm{Bi}_{4} \mathrm{Sr}_{3} \mathrm{Ca}_{3} \mathrm{Cu}_{y} \mathrm{O}_{x}(y=0-5)$ and $\mathrm{Bi}_{4} \mathrm{Sr}_{3} \mathrm{Ca}_{3}{ }_{z} \mathrm{Li}_{z} \mathrm{Cu}_{4} \mathrm{O}_{x}(z=0.1-1.0)$ semiconducting oxide glasses, J. Appl. Phys. {\bf 74}  (1993) 931. 

\bibitem{TaherAPA15} Y. B. Taher, A. Oueslati, N.K. Maaloul, K. Khirouni and M. Gargouri, Conductivity study and correlated barrier hopping (CBH) conduction mechanism in diphosphate compound, Appl. Phys. A {\bf 120} (2015) 1537.

\end{thebibliography}
\end{document}